\newif\iffigs\figstrue

\documentstyle[12pt]{article}

\iffigs
  \input epsf
\else
\fi

\batchmode
  \newfont{\footscrfont}{rsfs10}
  \newfont{\footbbbfont}{msbm10}
\errorstopmode

\newif\ifscrf\scrftrue
\ifx\footscrfont\nullfont
  \scrffalse
\fi

\newif\ifamsf\amsftrue
\ifx\footbbbfont\nullfont
  \amsffalse
\fi

\def\ppnumber{\vbox{\baselineskip14pt\hbox{CLNS-95/1366}
\hbox{hep-th/9510142}}}
\def\ppdate{October 1995}
\def\pplogo{\vbox{\kern-\headheight\kern -15pt
\halign{##&##\hfil\cr&{
\ppnumber}\cr\rule{0pt}{2.5ex}&\ppdate\cr}
}}

\makeatletter
\date{}
\def\dedicatory#1{\def\@date{\normalsize\it#1}}
\def\subjclass#1{\def\@thefnmark{}\@footnotetext{1991
    {\it Mathematics Subject Classification.} #1}}
\def\keywords#1{\def\@thefnmark{}\@footnotetext{
    {\it Key words and phrases.} #1}}

\def\ps@firstpage{\ps@empty \def\@oddhead{\hss\pplogo}%
  \let\@evenhead\@oddhead 
}
\def\maketitle{\par
 \begingroup
 \def\thefootnote{\fnsymbol{footnote}}
 \def\@makefnmark{\hbox
 to 0pt{$^{\@thefnmark}$\hss}}
 \if@twocolumn
 \twocolumn[\@maketitle]
 \else \newpage
 \global\@topnum\z@ \@maketitle \fi\thispagestyle{firstpage}\@thanks
 \endgroup
 \setcounter{footnote}{0}
 \let\maketitle\relax
 \let\@maketitle\relax
 \gdef\@thanks{}\gdef\@author{}\gdef\@title{}\let\thanks\relax}

\def\abstract{\if@twocolumn
\section*{Abstract}
\else \small
\begin{center}
{\bf ABSTRACT}
\end{center}
\quotation
\fi}

\def\thebibliography#1{\section*{References\@mkboth
 {REFERENCES}{REFERENCES}}\small\list
 {[\arabic{enumi}]}{\settowidth\labelwidth{[#1]}\leftmargin\labelwidth
 \advance\leftmargin\labelsep
 \usecounter{enumi}}
 \def\newblock{\hskip .11em plus .33em minus .07em}
 \sloppy\clubpenalty4000\widowpenalty4000
 \sfcode`\.=1000\relax}

\newif\iffn\fnfalse

\@ifundefined{reset@font}{\let\reset@font\empty}{} 
\long\def\@footnotetext#1{\insert\footins{\reset@font\footnotesize
    \interlinepenalty\interfootnotelinepenalty
    \splittopskip\footnotesep
    \splitmaxdepth \dp\strutbox \floatingpenalty \@MM
    \hsize\columnwidth \@parboxrestore
   \edef\@currentlabel{\csname p@footnote\endcsname\@thefnmark}\@makefntext
    {\rule{\z@}{\footnotesep}\ignorespaces
      \fntrue#1\fnfalse\strut}}}
\makeatother

\ifamsf
  \newfont{\bigbbbfont}{msbm10 scaled\magstep2}
  \newfont{\bbbfont}{msbm10 scaled\magstep1}  
  \newfont{\smallbbbfont}{msbm8}
  \newfont{\tinybbbfont}{msbm6}
  \newfont{\smallfootbbbfont}{msbm7}
  \newfont{\tinyfootbbbfont}{msbm5}
\fi

\ifscrf
  \newfont{\scrfont}{rsfs10 scaled\magstep1}  
  \newfont{\smallscrfont}{rsfs7}
  \newfont{\tinyscrfont}{rsfs7}
  \newfont{\smallfootscrfont}{rsfs7}
  \newfont{\tinyfootscrfont}{rsfs7}
\fi

\ifamsf
  \newcommand{\Bbb}[1]{\iffn
      \mathchoice{\mbox{\footbbbfont #1}}{\mbox{\footbbbfont #1}}
      {\mbox{\smallfootbbbfont #1}}{\mbox{\tinyfootbbbfont #1}}\else
      \mathchoice{\mbox{\bbbfont #1}}{\mbox{\bbbfont #1}}
      {\mbox{\smallbbbfont #1}}{\mbox{\tinybbbfont #1}}\fi}
\else
  \def\bigbbbfont{\bf}
  \def\Bbb{\bf}
\fi

\ifscrf
  \newcommand{\Scr}[1]{\iffn
    \mathchoice{\mbox{\footscrfont #1}}{\mbox{\footscrfont #1}}
    {\mbox{\smallfootscrfont #1}}{\mbox{\tinyfootscrfont #1}}\else
    \mathchoice{\mbox{\scrfont #1}}{\mbox{\scrfont #1}}
    {\mbox{\smallscrfont #1}}{\mbox{\tinyscrfont #1}}\fi}
\else
  \def\Scr{\cal}
\fi

\def\operatorname#1{\mathop{\rm #1}\nolimits}
\def\C{{\Bbb C}}

\def\P{{\Bbb P}}

\def\R{{\Bbb R}}
\def\Z{{\Bbb Z}}

\def\Hom{\operatorname{Hom}}
\def\Tors{\operatorname{Tors}}

\def\opeq#1{\advance\lineskip#1 \advance\baselineskip#1
	\advance\lineskiplimit#1}

\def\eqalign#1{\null\,\vcenter{\opeq{2.5\jot}\mathsurround=0pt
	\everycr={}\tabskip=0pt
	\halign{\strut\hfil$\displaystyle{##}$&$\displaystyle{{}##}$\hfil
	\crcr#1\crcr}}\,\null}

\def\CY{Calabi--Yau}

\def\cM{{\Scr M}}

\def\cD{{\Scr D}}

\def\cMc{{\hfuzz=100cm\hbox to 0pt{$\;\overline{\phantom{X}}$}\cM}}
\def\barcD{{\hfuzz=100cm\hbox to 0pt{$\;\overline{\phantom{X}}$}\cD}}

\def\ff#1#2{{\textstyle\frac{#1}{#2}}}

\ifamsf
  \def\ltimes{\mathbin{\mbox{\bbbfont\char"6E}}}
  
\else
  \def\ltimes{.}
  
\fi

\begin{document}
\setcounter{page}0
\title{\LARGE An N=2 Dual Pair\\and a Phase Transition\\[10mm]
}
\author{
Paul S. Aspinwall\\[0.7cm]
\normalsize F.R.~Newman Lab.~of Nuclear Studies,\\
\normalsize Cornell University,\\
\normalsize Ithaca, NY 14853\\[10mm]
}

{\hfuzz=10cm\maketitle}

\def\Large{\large}
\def\LARGE{\large\bf}

\vskip 1.5cm
\vskip 1cm

\begin{abstract}

We carefully analyze the $N=2$ dual pair of string theories in four
dimensions introduced by Ferrara, Harvey, Strominger and Vafa. The
analysis shows that a second discrete degree of freedom must be
switched on in addition to the known ``Wilson line'' to achieve a
non-perturbatively consistent theory. We also identify the phase
transition this model undergoes into another dual pair via a
process analogous to a conifold transition. This provides the
first known example of a phase transition which is understood from
both the type II and the heterotic string picture.

\end{abstract}

\vfil\break

\section{Introduction}		\label{s:intro}

Even though non-perturbative string theory has yet to be properly
defined we have gained some insight into its structure by using the
notion of duality. At this time no duality can be considered to be
rigorously proven simply because this would require a full definition
of non-perturbative string theory. What we can say however is that
some of the many proposed dualities do not contradict what we do know
of string theory and appear to be self-consistent. One might then take
these reasonably well-established dualities as a defining property of
string theory.

Perhaps the most powerful duality considered is that of string-string
duality \cite{W:dyn,HT:unity}. This states that the type IIA string
compactified on a K3 surface gives precisely the same six-dimensional
physics as the heterotic string compactified on a 4-torus. This
$N=2$ theory in six dimensions can then be compactified on a 2-torus
to yield an $N=4$ theory in four dimensions giving four dimensional
statements about duality as discussed in \cite{Duff:s,AM:Ud}.

Independently of this string-string duality in six dimensions, some
dual pairs for $N=2$ were proposed in four dimensions in
\cite{KV:N=2}. In these cases, the type IIA (or type IIB) string
compactified on some \CY\ manifold (or its mirror) is considered
to be equivalent to the heterotic string compactified on a conformal
field theory which is essentially derived from a K3 surface times a
2-torus. In \cite{KLM:K3f,VW:pairs} it was noted that the \CY\ manifolds
in question could be written as a fibration over a rational curve with
generic fibre a K3 surface whereas the K3$\times T^2$ case can be
written as a fibration over a rational curve with generic fibre
$T^4$. It is then tempting to try to use string-string duality in
six-dimensions fibre-wise to deduce these $N=2$ dual pairs in
four-dimensions. In \cite{VW:pairs} some preliminary observations were
made concerning this deduction but the full picture has not yet
emerged.

Another approach to obtaining more dual pairs in four dimensions is to
take orbifolds of the above $N=4$ case. As we do not yet know how to
deal with fixed points, it is best to start with free-actions on the
type II side. There are essentially two types of free quotient of
K3$\times T^2$. Firstly the action can be free on the torus and have
fixed points on the K3 surface. This leads to more $N=4$ theories
which were studied in \cite{CP:ao,me:dual,CL:KT/G}. Alternatively the
action is free on the K3 surface and has fixed points on $T^2$. This
is essentially unique and was first considered in
\cite{FHSV:N=2}. This example is the central object of study in this
paper. It gives
an $N=2$ theory in four dimensions but has many properties one would
associate with an $N=4$ theory. In this way it is probably the most
``tame'' $N=2$ theory in four dimensions.

The process of orbifolding to obtain the $N=4$ theories of
\cite{CP:ao,me:dual,CL:KT/G} was a straight-forward process. The
action of the quotienting group on the K3 surface and torus on the
type II side could be translated into the heterotic language to
produce the dual asymmetric orbifold. In this $N=4$ case the resulting
asymmetric orbifold is automatically modular invariant. We are not so
lucky with the $N=2$ case of \cite{FHSV:N=2} however and so the
orbifolding process must be more subtle if we are to obtain a dual
pair. The trick appears to be to use some degrees of freedom from the
R-R sector of the type II string. Such degrees of freedom are rather
trivial from the point of view of the conformal field theory on the
type II side but can effect the modular invariance properties of the
conformal field theory on the heterotic side. Thus there must be some
non-perturbative consistency requirement that we do not yet understand
on the type II side if duality is to work. In this paper we will give
a natural topological interpretation to the discrete degrees of
freedom which must be employed to satisfy this non-perturbative
consistency requirement. We will see that both the 1-form and the
3-form R-R fields in the type IIA string appear to play an important
r\^ole.

When one considers $N=2$ theories in four dimensions the possibility
of phase transitions must be considered. It was first noticed many
years ago \cite{Reid:never} that the moduli spaces of various \CY\
manifolds appeared to be ``connected'' to each other by going through
points where the 3-fold degenerates. This lead to the idea that
perhaps string theories corresponding to vacua given by different \CY\
manifolds could be deformed into each other \cite{GH:con}. The
simplest kind of degeneration considered for such a process is that of
a ``conifold''. The
immediate problem with this idea was that the conformal field theory
associated to the degenerate 3-fold was itself degenerate. Thus the
original idea was that string theory might ``tunnel'' between vacua.
More recently it was realized that although the conformal field
theory might be sick on the degenerate 3-fold, the complete
non-perturbative type II string was probably well-behaved
\cite{Str:con}. In fact, an example of a ``phase transition'' based on
a conifold concerning
type II string theories compactified on different \CY\ manifolds could
be understood in terms of the string theory as well as algebraic
geometry \cite{GMS:con}.

If there is a heterotic string theory that is dual to a given type II
string on a \CY\ manifold then one should be able to see this
transition in terms of the heterotic string too. For the conifold
transition studied in \cite{GMS:con} the heterotic dual is not yet
known. In \cite{KV:N=2} however, an example of a phase transition that
looked a lot like a conifold transition was given. Unfortunately the
type II dual for that example is not yet known. Recently some lists of
potential dual pairs were given in \cite{AFIQ:chains} where the
transitions are also understood from the heterotic side. In this paper we will
describe a case where both the type IIB string picture and the
heterotic string picture can be given.

It is perhaps worthwhile to spell out some nomenclature we will use
here as usage of the word ``conifold'' has varied in the past. The
term ``conifold'' was first introduced in \cite{CGH:con} to refer to
an algebraic variety with a finite number of nodes. A node is an
isolated singularity locally of the form
\begin{equation}
  w^2+x^2+y^2+z^2=0.
\end{equation}
Some later uses of the word ``conifold'' have been more general
than this. We will use the original definition.
Locally such a node can be resolved either by a deformation of complex
structure, which replaces the node by a single $S^3$, or by a ``small
resolution'' which replaces the node by an $S^2$. Within a conifold,
there can be an obstruction to the small resolution if we demand that
the resulting space is K\"ahler. If there is no such obstruction then
we have a conifold transition between two \CY\ manifolds. The
transition and degenerations considered in this paper are {\em not\/} of
this type. We will use the term ``extremal transition'' to refer to
the general process of a topology change due to being able to resolve
a singularity either by a deformation of complex structure or a
deformation of K\"ahler form (see, for example, \cite{Mor:look}). We
will also use the term ``phase
transition'' for this more general process. There are many extremal
transitions which are not conifold transitions. An example based on a
orbifold was studied in \cite{VW:tor,AMG:stab}.

The original proposal in \cite{Reid:never} was that {\em all\/} \CY\ manifolds
should somehow be connected together by extremal transitions. It was
stated in rather modest terms however. Firstly only simply-connected
manifolds were considered and secondly non-K\"ahler manifolds were
allowed as intermediate states. It was seen in \cite{GH:con}
that a surprisingly large class of simply-connected manifolds could be
connected
without the need for non-K\"ahler manifolds. It was suggested in
\cite{GMS:con} that perhaps all $N=2$ theories in four dimensions
might be connected by phase transitions. This leaves open the issue of
$\pi_1$ however. For reasons we will give later, one might suspect
that $\pi_1$ should not change. We will show that in our example
$\pi_1$ can indeed change and so this potential obstacle for
connecting all \CY\ manifolds together is removed.

The paper is divided into two parts --- one dealing with the analysis
of the discrete degrees of freedom of the type II string and one
dealing with the phase transition. The two parts can be read
independently of each other but it is worth emphasizing that they
should be considered to be intimately linked as they ultimately both
analyze exactly the same vector within the lattice $\Gamma^{6,22}$.
In section \ref{s:orb} we will analyze carefully the complete moduli
space of type II strings compactified on a space in the form of $Y/G$,
where $G$ acts freely. This will allow us in section \ref{s:het} to
identify exactly which type IIA string is dual to the heterotic string
in the dual pair we consider. In section \ref{s:deg} we will examine
closely a point in the moduli space of the type IIA and type IIB
string where a rather curious degeneration occurs. In section
\ref{s:pht} we show how this leads to the phase transition already
understood from the heterotic side from \cite{FHSV:N=2}. Finally in
section \ref{s:conc} we make some concluding remarks.


\section{Orbifolds Without Fixed Points} \label{s:orb}

We wish to consider orbifolds in the form $X=Y/G$ where $Y$ is a
smooth manifold and $G$ is a discrete group that acts freely. Thus $X$
is also a smooth manifold.
Of particular interest to us is the detailed form of the moduli space
of string theories on $X$. This comes from the singular cohomology of $X$
as we now discuss. The 10-dimensional form of a string theory contains
fields described by $p$-forms for various values of $p$. That is, the
action will have $(p+1)$-form field-strengths. Such fields in the
10-dimensional picture can lead to moduli in the lower-dimensional
theory upon compactification on $X$. The general picture is that one can
``integrate'' the $p$-form over some $p$-cycle within $X$. Thus the
number of moduli obtained per $p$-form would appear to be $b_p=\dim
H_p(X)$. (In general more moduli can appear from duality
transformations but we need not concern ourselves with those here.)

As well as the dimension of the moduli space arising from the
$p$-forms in ten-dimensional space-time we would also like to know the
precise form of this moduli space. Given the lack of a precise
definition of string theory we cannot really establish anything
rigorously here but we can make some reasonable guesses. Firstly we
will assert some
periodicity on these moduli. This amounts to saying that shifting the
moduli by an element of $H^p(X,\Z)$ results in a shift in the action
by $2\pi i$ and hence has no effect on the resulting physics. This
periodicity can be viewed as arising from the ``quantization'' of
soliton charges \cite{HT:unity} and was shown to follow directly from
duality in the case of K3$\times T^2$ in \cite{AM:Ud}. Thus far we
appear to be saying that the moduli space is
$H^p(X,\R)/H^p(X,\Z)$. There can also be extra discrete degrees of freedom
coming from torsion cycles within $X$. That is, we may have a singular
cycle, $C$, which is homologically non-trivial but $n[C]=0$ for some
integer $n$. We can consistently attach phases to such cycles to give
discrete degrees of freedom to the moduli. This extra degree of
freedom is not particularly new as in the case $p=1$ they correspond
to nothing more than ``Wilson lines''. The case $p=2$ has already been
considered in string theory for the $B$-field in the context of
``discrete torsion'' \cite{Vafa:tor}.

All said, we claim that each $p$-form in the ten-dimensional picture
contributes a factor
\begin{equation}
\eqalign{
  \cM_p &= U(1)^{b_p} \times \Tors(H_p(X))\cr
  &= U(1)^{b_p} \times \Tors(H^{p+1}(X))\cr
  &= H^p(X,U(1))\cr
  &= \Hom(H_p(X),U(1))\cr}	\label{eq:Mp}
\end{equation}
to the moduli space, where $\Tors$ denotes the torsion part. See
\cite{BT:} for the methods used to obtain the various equalities in
(\ref{eq:Mp}). Our notation will always assume (co)homology to be over
the integers unless otherwise specified.
In the case of $p=1$ we have $\cM_1 = \Hom(H_1(X),U(1)) =
\Hom(\pi_1(X),U(1))$. This is precisely the moduli space of flat
$U(1)$-bundles on $X$. For $p>1$ we are making a similar statement for
``currents'' arising from $(p-1)$-branes. In particular, as stated above,
we have already encountered $\cM_2$ in string theory
\cite{Vafa:tor,AM:suff}.

Now we want to determine the moduli space for the orbifold $X=Y/G$.
Locally it is certainly the case that the moduli of
$X$ are the $G$-invariant moduli of $Y$. Thus, we first need to
establish the moduli space of $Y$. Then we should consider the
possibility that
there may be homology cycles in $X$ that did not descend
from $Y$. That is, the quotient map $f:Y\to X$ yields
\begin{equation}
  f_*:H_*(Y)^G \to H_*(X)
\end{equation}
which is injective but not an isomorphism. $G$ as a superscript denotes the
$G$-invariant part. The cokernel of this map gives
us precisely the extra discrete degrees of freedom present in the
orbifold. In this paper we will work with cohomology as well as
homology. The reader is referred to \cite{BT:} for the detailed
relationship between these groups.

Our main weapon for comparing the cohomology of $X$ and $Y$ is the
{\em Hochschild-Serre spectral sequence} \cite{Milne:}. That is, there
is a spectral sequence with second stage
\begin{equation}
  E_2^{p,q} = H^p(G,H^q(Y)).
\end{equation}
This converges to $E_\infty^{p,q}$ where
\begin{equation}
  H^n(X) = \bigoplus_{p+q=n} E_\infty^{p,q}.
\end{equation}

Detailed analysis of this spectral sequence can be rather messy in
general but we can make some simple statements. Firstly the term
$E_2^{p+1,0}=H^{p+1}(G)$ is simple group cohomology (that is, of the
form $H^*(G,\Z)$ where $G$ acts trivially on $\Z$). Unless killed by
earlier stages of the spectral sequence this will have either of two
effects on $H^n(X)$. Firstly it can remove some of $E_2^{0,p}\cong
H^p(Y)^G$ from
$H^{p}(X)$. Secondly it can contribute torsion to $H^{p+1}(X)$. Either
of these effects will increase the discrete degrees of freedom in
$\cM_p$ relative to the moduli space of $Y$.
If we assume $Y$ is
simply connected and $G$ is finite then for the case
$\cM_1$ these new discrete
degrees of freedom are in fact given precisely by $H^2(G)$, which is the
abelianization of $G$, and for the case
$\cM_2$ we have $H^3(G)\cong H^2(G,U(1))$. This
simplification explains why we have not had to suffer the complexities
of spectral sequences in the past. For $p=1$, the new degrees of
freedom, ``Wilson Lines'' were given simply by the quotienting group
itself and for $p=2$ we only had to worry about basic group
cohomology to find the ``discrete torsion''. In the cases $p>2$ life
appears to demand the Hochschild-Serre spectral sequence in its full
glory to understand the effect of orbifolding.

In general other entries
in $E_2^{p,q}$ will also effect these discrete degrees of freedom. In
particular, one of the contributions can come from the fact that the
moduli space of $G$-invariant theories on $Y$ is not connected even
though the complete moduli space of theories on $Y$ is connected.

Let $Y$ be a product of a K3 surface and a 2-torus as in
\cite{FHSV:N=2}. $Y$ admits a
symmetry $G\cong\Z_2$. The generator, $g$, of $G$ acts freely on the
K3 surface. The quotient K3$/G$ is known as the ``Enriques Surface''
and is not a \CY\ surface. In particular, if $\omega$ is a (2,0)-form
on the K3 surface then
\begin{equation}
  g:\omega\to-\omega.	\label{eq:Enri}
\end{equation}
If the 2-torus is
represented by periodic identification of the 2-plane with coordinates
$(x,y)$ then $g:(x,y)\to(-x,-y)$. Thus $g$ has 4 fixed points on
$T^2$. The complete action of $g$ is free as required.

The topology of $X$ is very rich and is one of the reasons why a full
understanding of string duality on this manifold would provide great
insight into the general case. The global holonomy of $X$ is
$SU(2)\times\Z_2$. Thus, although $Y$ is a \CY\ manifold it does not
have the generic holonomy $SU(3)$. This leads to some simplifications
such as no world-sheet instanton corrections to the type II string.

Next consider the fundamental group. If $\overline{X}$ is the
simply-connected covering space of $X$ then
$\overline{X}/\pi_1(X)\cong X$. Thus
\begin{equation}
  \pi_1(X) \cong \Z_2 \ltimes (\Z)^2,
\end{equation}
where the $\Z_2$ acts on the lattice $(\Z)^2$ as $g$ acted on the
torus.

\begin{figure}
\setlength{\unitlength}{0.9mm}%
$$\begin{picture}(130,90)(0,0)
\thinlines
\put(0,0){\line(1,0){130}}
\put(0,0){\line(0,1){90}}
\put(0,0){\makebox(25,20){$\Z$}}
\put(25,0){\makebox(25,20){$0$}}
\put(50,0){\makebox(25,20){$\Z_2$}}
\put(75,0){\makebox(25,20){$0$}}
\put(100,0){\makebox(25,20){$\Z_2$}}
\put(0,20){\makebox(25,20){$0$}}
\put(25,20){\makebox(25,20){$(\Z_2)^2$}}
\put(50,20){\makebox(25,20){$0$}}
\put(75,20){\makebox(25,20){$(\Z_2)^2$}}
\put(100,20){\makebox(25,20){$0$}}
\put(0,40){\makebox(25,20){${\displaystyle H^2(Y)^G\atop
		\displaystyle\cong(\Z)^{11}}$}}
\put(25,40){\makebox(25,20){$(\Z_2)^2$}}
\put(50,40){\makebox(25,20){$\Z_2$}}
\put(75,40){\makebox(25,20){$(\Z_2)^2$}}
\put(100,40){\makebox(25,20){$\Z_2$}}
\put(0,60){\makebox(25,20){${\displaystyle H^3(Y)^G\atop
		\displaystyle\cong(\Z)^{24}}$}}
\put(25,60){\makebox(25,20){$0$}}
\put(50,60){\makebox(25,20){$(\Z_2)^4$}}
\put(75,60){\makebox(25,20){$0$}}
\put(100,60){\makebox(25,20){$(\Z_2)^4$}}
\put(43,46.5){\vector(2,-1){65}}
\put(18,66.5){\vector(2,-1){65}}
\put(62.5,-3){\makebox(0,0){$p$}}
\put(-3,50){\makebox(0,0){$q$}}
\put(62,10){\circle{10}}
\put(62,50){\circle{10}}
\end{picture}$$
\caption{Hochschild-Serre spectral sequence term for
          $E_2^{p,q}$.}	  \label{fig:SS}
\end{figure}
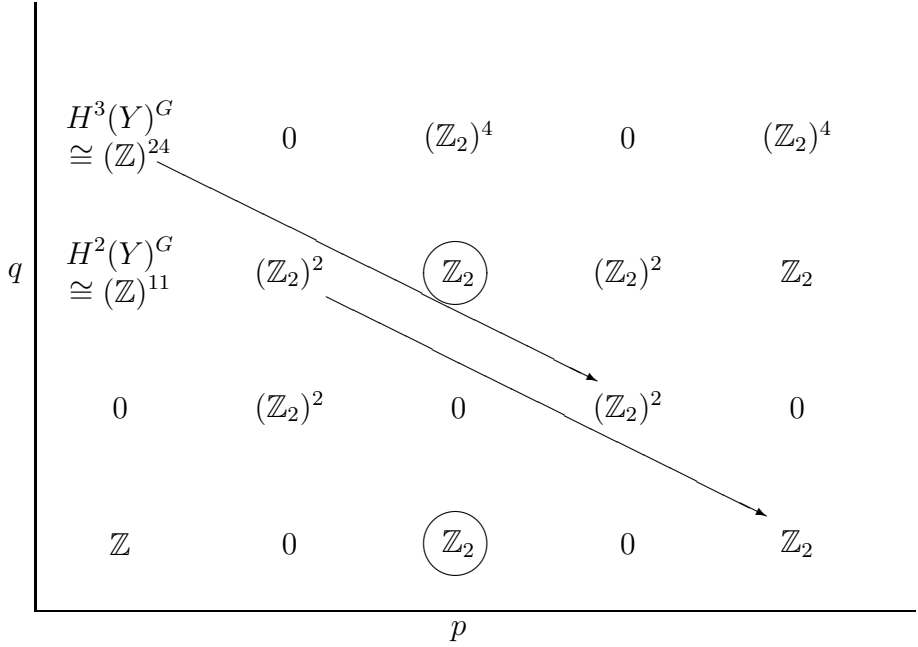

We can now give the cohomology for the case in question from the above
discussion. In terms of homology, which is clearer for our purposes,
\begin{equation}
  \eqalign{
     H_0(X) &= \Z\cr
     H_1(X) &= (\Z_2)^3\cr
     H_2(X) &= (\Z)^{11}\times\Z_2, \quad H_2(X)/(f_*H_2(Y)^G) = \Z_2\cr
     H_3(X) &= (\Z)^{24}\times\Z_2, \quad H_3(X)/(f_*H_3(Y)^G) = (\Z_2)^3\cr
     H_4(X) &= (\Z)^{11}\times(\Z_2)^3\cr
     H_5(X) &= 0\cr
     H_6(X) &= \Z.\cr}	\label{eq:HX}
\end{equation}
This is obtained from the spectral sequence shown in figure
\ref{fig:SS}. To compute the exact progression of this spectral
sequence we need to compare this spectral sequence to that for the
Enriques surface as quotient of a K3 surface.
We also need to know that $X$ can be written as an
elliptic fibration over an Enriques surface and that this bundle has a
section. One can then show that $H^3(X)$ has torsion as given in
(\ref{eq:HX}) and that the two $d_3$ maps shown in figure \ref{fig:SS}
are surjective. Note that we recover the result $h^{1,1}(X)=h^{2,1}(X)=11$.

Now let us consider compactifying the type IIA superstring on the \CY\
manifold $X$. In order to analyze the moduli space we first need the
(connected part of) the moduli space
of a IIA string on $Y$ that admits $G$ as a symmetry. This is readily
shown to be
\cite{FHSV:N=2}
\begin{equation}
  \cM_Y^G \cong \cM_Y^V \times \cM_Y^H,	\label{eq:MY}
\end{equation}
where
\begin{equation}
  \eqalign{\cM_Y^V &\cong Sl(2,\Z)\backslash Sl(2)/U(1) \times
O(\Gamma^{2,10})\backslash
O(2,10)/(O(2)\times O(10))\cr
  \cM_Y^H &\cong O(\Gamma^{4,12})\backslash
O(4,12)/(O(4)\times O(12)),\cr}	\label{eq:MYs}
\end{equation}
where $\Gamma^{p,q}$ is a lattice with intersection form signature
$(p,q)$ and $O(\Gamma^{p,q})$ is its group of rotational
isometries. Note that
neither $\Gamma^{2,10}$ nor $\Gamma^{4,12}$ in (\ref{eq:MYs}) are
self-dual.

{}From (\ref{eq:HX}) we can now find the extra discrete degrees of
freedom exhibited by the type IIA string on $X$. Firstly the NS-NS
sector of the type IIA string has a 2-form. This gives a $\Z_2$-valued
parameter. The NS-NS degrees of freedom should be visible from the
conformal field theory. This is not really ``discrete torsion'' in the
sense of \cite{Vafa:tor} as $H^3(\Z_2)=0$ but rather has its origins
in the fact that the moduli space of $G$-invariant theories on $Y$ is
not connected.
We will not discuss this further although it
would be interesting to analyze this in terms of mirror symmetry and
the consequences for the heterotic dual picture.

In the R-R sector we have a 1-form and a 3-form. First take the
1-form. The first homology of $X$ is entirely torsion in the form
$(\Z_2)^3$. This gives us effectively a choice of 7 non-trivial Wilson
lines. Two of the $\Z_2$'s may be considered to be remnants of the
first homology of the $T^2$. The other $\Z_2$ corresponds to usual
picture of lines in $Y$ whose ends are identified by $G$ descending to
closed loops in $X$.

The 3-form also contributes non-trivial choices. In this case the
extra homology is of the form $(\Z_2)^3$ where one of these factors
appears as torsion in $H_3$. Again two of these factors are remnants
of the 1-cycles on the torus. The torsion term comes from $E_2^{2,2}
= H^2(Y)^G/((1+g)H^2(Y))$. This also has origins in the torus --- it
arises from the class in $H^2(Y)$ which is dual to the homology class
of the torus itself.


\section{The Heterotic Dual}	\label{s:het}

If we accept that the type IIA string compactified on a K3 surface is
dual to the heterotic string on $T^4$ then it follows that the type
IIA string on $Y$ is dual to a heterotic string on $T^6$. One can then
try to orbifold this dual pair by our $\Z_2$-action to form another
dual pair with $N=2$ supersymmetry in four dimensions. Making this
orbifolding process non-perturbatively correct is not a simple matter
however \cite{FHSV:N=2,VW:pairs}.

First let us interpret the moduli space given in (\ref{eq:MY}). The
moduli space of type IIA strings on $Y$ is given by \cite{HT:unity,AM:Ud}
\begin{equation}
  \cM_Y = O(\Gamma^{6,22})\backslash O(6,22)/(O(6)\times O(22))
	\times Sl(2,\Z)\backslash Sl(2)/U(1).    \label{eq:MY0}
\end{equation}
In this
case $\Gamma^{6,22}$ is an even self-dual lattice that can be
identified as $\Gamma^{6,22}=\Gamma^{4,20}\oplus\Gamma^{2,2}$, where
$\Gamma^{4,20}\cong H^*({\rm K3})$ \cite{AM:K3p} and $\Gamma^{2,2}$
is the Narain lattice $H^1(T^2)\oplus H_1(T^2)$ as in \cite{N:torus}.

Given that $\Gamma^{6,22}$ can also be identified as the Narain
lattice for compactification of the heterotic string we now have an
idea of the identification between these dual pairs. In particular, as
we know how $G$ acts on the cohomology of $Y$, we can almost see how it acts
on the Narain lattice. Unfortunately our knowledge is not
complete. What we find by this route is only how $G$ acts by {\em
rotations\/} on the Narain lattice. We also need to know about
translations.

In order to understand the translations we look at the gauge group of
the string theory. On the heterotic side, for a generic torus, the
gauge group arises from the metric \`a la Kaluza-Klein from
isometries of the torus. This corresponds to the translations of the
Narain lattice and the gauge group is generically $U(1)^{28}$. How
does this $U(1)^{28}$ arise on the type IIA side and how does this fit
into the type II interpretation of $\Gamma^{6,22}$?

Firstly we have a metric in the ten-dimensional theory which produces
gauge fields from isometries, i.e., $H^1(Y)$. Secondly we have a
2-form which compactifies on 1-cycles $H_1(Y)$ to give gauge
fields. These two contributions obviously come from the $\Gamma^{2,2}$
of $T^2$.
Next we have a 1-form in the ten-dimensional theory which descends
trivially down into a $U(1)$ gauge group upon compactification. It seems
natural to identify this with the $H^0({\rm K3})$ direction. We also
have a 3-form which compactifies on 2-cycles to give gauge fields. 22
of these 2-cycles come from $H_2({\rm K3})\cong H^2({\rm K3})$. The
last gauge field comes from the 2-cycle given by $T^2$ itself. Within
$Y$ this homology element maps to $H^4({\rm K3})$ under the
isomorphism $H_2(Y) \cong H^4(Y)$. This completes our identification of
all 28 gauge fields with shifts in $\Gamma^{6,22}$.

If we ignore translations and take the asymmetric orbifold of the
heterotic string by the rotation of $\Gamma^{6,22}$ induced by $G$, it
was observed in \cite{FHSV:N=2} that one could not obtain a modular
invariant theory because of level-matching problems
\cite{NSV:asym}. If $g$ acts by both a rotation and a shift
simultaneously however, one can satisfy level-matching.

The shift used in \cite{FHSV:N=2} was uniquely identified as a
half-shift along the lattice vector of length (squared) $-2$ contained in the
lattice generated by $H^0({\rm K3})$ and $H^4({\rm K3})$. That is,
both $H^0({\rm K3})$ and $H^4({\rm K3})$ must be shifted as each of
these directions alone has length zero.
The ability to shift by this half-vector of the lattice must manifest
itself as one of the discrete degrees of freedom of the orbifold
theory on $X$. Let us try to find exactly which degree of freedom it
should be.

Since the $H^0({\rm K3})$ direction comes from the 1-form in
ten-dimensions, this must be associated with an element of
$H_1(X)$. The new element of $H_1(X)$ introduced in the quotienting
process was the ``Wilson line'' joining points in $Y$ identified by
$G$. Thus, this shift in $H^0({\rm K3})$ corresponds to ``switching
on'' this ``Wilson line'' degree of freedom in the orbifold $X$. This
was also observed in \cite{FHSV:N=2}.

The $H^4({\rm K3})$ direction comes from the 3-form in ten-dimensions
compactified over the $T^2\subset Y$. As explained above, a new
element in $H_3(X)$ arises from this torus from the $E_2^{2,2}$ term
in the spectral sequence. It would thus appear that we also have to
switch on this discrete degree of freedom too to obtain a truly
consistent non-perturbative string theory.

To summarize, the R-R fields in the 10-dimensional type II string
theory give discrete parameters in the compactified theory. Modular
invariance of the dual heterotic string theory demands that some of
these parameters are ``switched on''. One is associated with 1-cycles
on $X$ and one is associated with 3-cycles on $X$. Both of these
cycles are torsion elements of the homology of $X$. The origins of
these discrete degrees of freedom from the spectral sequence are
circled in figure \ref{fig:SS}.


\section{Degenerations}	\label{s:deg}

Consider the moduli space of type II string theories on K3$\times T^2$. Can we
undergo a phase transition to another class of theories? That is, are
there any extremal transitions of K3$\times T^2$? The answer is
no. To undergo an extremal transition one first goes to the
boundary of moduli space a finite distance away where something
degenerates. This we can do by letting the K3 surface acquire an
orbifold singularity. We then try to resolve this singularity in a way
inequivalent to how we reached the degeneration. This we cannot do ---
any attempt to resolve the orbifold
singularity will return us to K3$\times T^2$. Thus, type II strings on
K3$\times T^2$ appear to be an isolated island in the grand moduli
space of all theories. The same is also true of each of the other $N=4$
quotients of \cite{CP:ao,me:dual,CL:KT/G}.

Can we expect anything different for $X=({\rm K3}\times T^2)/\Z_2$
then? In \cite{FHSV:N=2} it was argued from the heterotic side that
there was a point in a given connected component of the moduli
space of these theories where a phase transition was possible. In this
section we will argue the same from the type II side.

How can $X$ acquire a singularity? There are two ways. Firstly the
covering space $Y$ may be singular. From what we said above however,
such a degeneration cannot be interesting from the point of view of
phase transitions as all we can do to move away from such a point is
to go back into the moduli space from whence we came. Alternatively
the quotienting process by $G$ may degenerate. That is, $G$ is meant
to act freely but at the boundary of the moduli space it can acquire
fixed points. This latter case offers a doorway into other moduli
spaces as we shall see.

The moduli space of type II strings on $Y$ was given in
(\ref{eq:MY0}). The first factor is the only one of interest to us and
may be regarded as
quotient of the Grassmanian of space-like 6-planes in $\R^{6,22}$ by
the isometry group of the even self-dual lattice $\Gamma^{6,22}\subset
\R^{6,22}$.
The connected
component of the moduli space of $G$-invariant theories given in
(\ref{eq:MY}) is obtained easily from this by analyzing how $G$ acts
on the lattice $\Gamma^{6,22}$ and the 6-plane. From our knowledge of
how to interpret $\Gamma^{6,22}$ in terms of $Y$ from section
\ref{s:het} we can decompose
$\Gamma^{6,22}$ as
\begin{equation}
  \Gamma^{6,22} = \Gamma_1^{1,9}\oplus \Gamma_2^{1,9}
  \oplus \Gamma_1^{1,1}\oplus\Gamma_2^{1,1}\oplus\Gamma^{2,2},
\end{equation}
where each of the above terms on the right are even self-dual,
and then identify the action of $G$ as generated by
\begin{equation}
g:(\Gamma_1^{1,9},\Gamma_2^{1,9},\Gamma_1^{1,1},\Gamma_2^{1,1},
\Gamma^{2,2})\mapsto (\Gamma_2^{1,9},\Gamma_1^{1,9},\Gamma_1^{1,1},
-\Gamma_2^{1,1},-\Gamma^{2,2}).   \label{eq:gG}
\end{equation}

The result given in
(\ref{eq:MYs}) is derived given that $\Gamma^{2,10}\subset
\Gamma^{6,22}$ is the $G$-invariant part of the lattice and
$\Gamma^{4,12}\subset\Gamma^{6,22}$ is the set of lattice vectors
orthogonal to $\Gamma^{2,10}$. The 6-plane itself, which we will call
$\Pi_6$, must be the product
of a 2-plane, $\Pi_2$, which is invariant under $G$ and a 4-plane,
$\Pi_4$, such that any vector in $\Pi_4$ is inverted by $g$.
The result is that the moduli space is a quotient of the product of
the Grassmanian of space-like 2-planes in $\R^{2,10}\supset\Gamma^{2,10}$
and the Grassmanian of space-like 4-planes in
$\R^{4,12}\supset\Gamma^{4,12}$. The former is a factor in the moduli
space generated by the scalars coming from vector multiplets
and the latter forms the moduli space generated by the
hypermultiplets.

The lattice $\Gamma^{2,10}$ is naturally interpreted as the lattice of
invariant cohomology on the original K3 surface. This lattice contains
the Picard lattice of the K3 surface. The space, $\cM_Y^V$ of (\ref{eq:MYs}),
associated with this is therefore identified with moduli space of
K\"ahler forms and $B$-fields on $X$. Note that the moduli space of
K\"ahler forms and $B$-fields on the original $T^2$ gives a separate
$Sl(2,\Z)\backslash Sl(2)/U(1)$ factor. The space $\cM_Y^H$ comes from the
moduli space of complex structures on $X$, the R-R moduli and the
axion-dilaton system. $\cM_Y^H$ has a quaternionic structure.

For reasons explained in \cite{W:dyn,me:enhg} we expect interesting
things to happen in a type II string theory on $Y$ whenever the
space-like 6-plane is orthogonal to elements of $\Gamma^{6,22}$ of
length $-2$. In particular the set of such elements should give the
root diagram of a semi-simple group which will be the non-abelian part
of the gauge group of the string theory. The underlying conformal
field theory is expected to degenerate at such a point
\cite{me:enhg,W:dyn2}. This is therefore a prime candidate for a point
at the edge of moduli space where we might expect a phase transition
to live.

Consider an element, $e_1\in\Gamma_1^{1,9}$, of (\ref{eq:gG}) of
length $-2$. Let us try to make this orthogonal to the 6-plane,
$\Pi_6$. Let $e_2=g(e_1)$, that is $e_2$ is a vector of length $-2$ in
$\Gamma_2^{1,9}$. We see that $e_1+e_2\in\Gamma^{2,10}$ and
$e_1-e_2\in\Gamma^{4,12}$. If we arrange $\Pi_2$ so that it is
orthogonal to $e_1+e_2$ and $\Pi_4$ so that it is orthogonal to
$e_1-e_2$ then we will have $\Pi_6$ orthogonal to $e_1$ as desired. We
also necessarily have $e_2$ orthogonal to $\Pi_6$ at the same time. We
can identify these two vectors with $S^2$'s within the K3
surface in $Y$ as explained in \cite{me:enhg}. When we move $\Pi_2$
and $\Pi_4$ to be orthogonal to these vectors we cause these spheres
to shrink down to zero size. We also fix the $B$-field in the
direction associated with the homology classes of these spheres {\em
and\/} we have also fixed the R-R moduli to have zero values in these
directions.

Thus the situation is very similar to that of type IIA strings on a K3
surface. At particular places in the moduli space we obtain an
enhanced gauge symmetry. This is associated with 2-spheres shrinking
down to zero size. For the IIA string on a K3 surface we had to impose
a vanishing condition on the $B$-field and in the above case we also
have to impose a condition on the R-R moduli.

The effect of quotienting by $G$ is to identify the two $S^2$'s given
by $e_1$ and $e_2$. Thus we only get generically a single $SU(2)$ as the gauge
group on $X$. The conformal field theory associated to the heterotic
string picture of this theory will have the associated affine algebra
at level 2 however. By applying the same procedure to further
length $-2$ vectors we may enhance the gauge group beyond $SU(2)$ in the
usual way.

What we have just described is the process of $Y$ becoming singular
because the covering space $X$ has acquired singularities. Not
surprisingly this turns out to locally look like a dimensional
reduction of the type IIA string on a K3 surface case. That is, in
addition to the vector multiplet becoming massless we also have a
hypermultiplet in the same representation of the gauge group
\cite{FHSV:N=2}.
We can also see this directly from the heterotic picture. In this case
we obtain massless states due to winding/momentum modes around $\pm e_1$
and $\pm e_2$. The fact that we can form both invariant and
anti-invariant vectors from $e_1$ and $e_2$ shows that we get
a hypermultiplet and a vector multiplet from each.
This is like the $N=4$ Yang-Mills theory of
\cite{SW:II}. There are no phase transitions associated to this as we
expected.

In order to achieve this degeneration we needed to choose a special
point within $\cM_V$ and $\cM_H$ at the same time. That is, both the
vector multiplet and hypermultiplet moduli had to be tuned
simultaneously. If we want to obtain a degeneration by tuning just one
of the sets of moduli we need to look for length $-2$ vectors within
$\Gamma^{2,10}$ or $\Gamma^{4,12}$. Let us focus on the vector
multiplet moduli space, i.e., look for a vector of length $-2$ within
$\Gamma^{2,10}$. There are only two such vectors which we denote $e_F$
and $-e_F$. Thus, if $\Pi_2$ is perpendicular to
$e_F$ then we have an enhanced gauge group $SU(2)$ again.
What is the geometrical interpretation of this?

We have identified $\Gamma^{2,10}$ as the invariant part of the total
cohomology of the K3 surface. The only vectors of length $-2$ in this
lattice are $\pm(e_0-e_4)$ where $e_0$ generates $H^0({\rm K3})$ and
$e_4$ generates $H^4({\rm K3})$ --- i.e., the same direction as we
considered in section \ref{s:het}! When we use the results of
\cite{AM:K3p,AM:K3m} to interpret a point in the moduli space where
$\Pi_2$ is orthogonal to $e_f$, we deduce that the K3 surface
has volume of order the Planck scale. That is, it is the effects of
quantum geometry on the K3 surface that induces the enhanced gauge
group. We can usually understand quantum geometry in terms of
classical geometry by using mirror symmetry and this is no exception.

The mirror of a type IIA string theory compactified on $X\cong ({\rm
K3}\times T^2)/G$ is a type
IIB string compactified on another space $X^\prime\cong({\rm K3}\times
T^2)/G$. $X$ and $X^\prime$ are a mirror pair and are topologically
the same. The moduli of $X$ and $X^\prime$ are not generically the same
however. Let us analyze the type IIB string compactified on $X^\prime$.

To solve a notational problem, rather than considering a
type IIB theory compactified on $X^\prime$, from now on we will
consider the type IIB theory to be compactified on $X$. This means
that $\Gamma^{2,10}$ is now {\em anti}-invariant under $g$ and
$\Gamma^{4,12}$ is invariant in the type IIB context. In particular,
$\cM_Y^V$ is now the moduli space of
complex structures on $X$ and $\cM_Y^H$ is the moduli space of
``quaternionified'' K\"ahler forms. Thus, rather than requiring $X$ to
be a particular size to get our $SU(2)$ gauge group in the type IIA
case, we can ask that $X$ takes a particular complex structure in the
type IIB case.

For type IIB strings, the lattice $\Gamma^{2,10}$ is now identified
with the lattice {\em orthogonal\/} to the invariant sublattice of
$H^*({\rm K3})$. This is a subspace of $H^2({\rm K3})$. We can
view $\Pi_2$ as the 2-plane in $H^2({\rm K3},\R)$ which is spanned by
the real and imaginary parts
of the holomorphic 2-form, $\omega$, on the K3 surface.

\iffigs
\begin{figure}
  \centerline{\epsfxsize=12.5cm\epsfbox{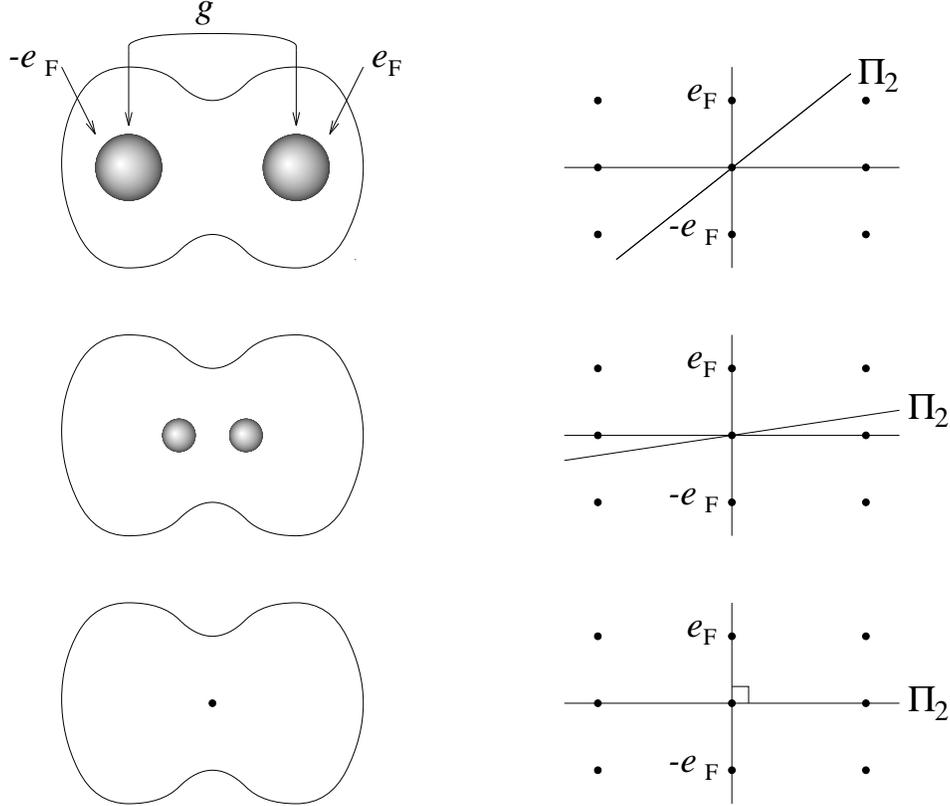}}
  \caption{The degeneration of the K3 surface for $e_F$.}
  \label{fig:a}
\end{figure}
\fi

Let us suppose first that $\Pi_2$ is not perpendicular to $e_F$.
We have an $S^2$ within the K3 surface whose homology class
is given by $e_F$. As $g$ acts freely on the K3 surface there must be
another sphere in the K3 surface which is the image of the first
sphere under $g$ and whose homology class is $-e_F$. See figure
\ref{fig:a} for this configuration. Now rotate $\Pi_2$ to be
perpendicular to $e_F$. Now these spheres should become rational curves
as they are dual to (1,1)-forms. This cannot be however since this
would imply that the curves given by $e_F$ and $-e_F$ are both
``effective'' which is
impossible (see, for example, \cite{BPV:}). Thus something must have
gone wrong. That is to say, we
simply cannot build an Enriques surface as the free quotient of a K3
surface if the K3 surface has this special complex structure. Nothing
has gone wrong with the K3 surface itself so far as its complex
structure is concerned so it must be the quotienting process
which has degenerated.

The way to picture the situation is shown in figure \ref{fig:a}. When
$e_F$ is orthogonal to $\Pi_2$, it becomes a rational curve at which
point its area is given by integrating the K\"ahler form over it. In
this case, the K\"ahler form is invariant under $G$ and so must be
orthogonal to $e_F$. Thus the areas of both spheres are zero. This is
not all that happens however. In order to resolve the paradox above
these spheres must become a fixed point of the quotienting. That is to
say, as we rotate $\Pi_2$ to become perpendicular to
$e_F$, the two spheres in the K3 surface both shrink down to zero size
and coalesce. Taking the orbifold of this K3 surface by this $\Z_2$
action will result in a degenerate Enriques surface. This surface is quite
interesting and has been
analyzed in many places such as \cite{Mor:Enr,Shah:Enr}. In our case,
the threefold $X$
will degenerate to singular object which we call $X^\sharp$.

Now let us review the heterotic side of this story given in
\cite{FHSV:N=2}. The winding/momentum modes around $\pm e_F$ again
give massless modes. As $\pm e_F$ are invariant under $g$ we get only
vector multiplets from these lattice
elements. The gauge group has been enhanced to $SU(2)$. In addition to
these massless states we obtain more from the twisted sector. The
energy of a $g$-twisted left-moving string of length $-l^2$ with no
oscillators is
\begin{equation}
\eqalign{E_L^g &= \ff14\sum_{i=1}^{22}g_i(1-g_i) -1 + \ff12l^2\cr
  &= \ff12l^2-\ff14,\cr}
\end{equation}
where the eigenvalues of the rotation of the lattice are given by
$\exp(2\pi i.g_i)$.

Recall that in section \ref{s:het} we made $g$ act by a translation on
the lattice $\Gamma^{6,22}$ by $\ff12e_F$. Thus, the shortest strings
in the twisted sector are those given by the vectors $\pm\ff12e_F$ and
these have zero energy. That is, they are massless states. This zero
energy-level was precisely the reason why this translation was
introduced in the first place --- in order to satisfy
level-matching. Under $g$, these lattice vectors are anti-invariant
and so these new states in the heterotic string are hypermultiplets. They
also have half the $U(1)$-charge of the vector multiplets which became
massless to give the $SU(2)$ gauge group. It follows that they are
${\bf 2}$'s of $SU(2)$. Following methods from \cite{NSV:asym} one can
count the degeneracy and finds one gets 4 states. All counted therefore,
at this point corresponding to the type IIB string on $X^\sharp$ we
obtain an enhanced gauge group of $SU(2)$ with four hypermultiplets in
the fundamental representation becoming massless. This situation
was considered in \cite{SW:II} where it played a special r\^ole due to
the fact it is conformally invariant. Conformal invariance was
important for consistency of the duality picture too as explained in
\cite{FHSV:N=2}.

We now see a peculiar connection between this section and section
\ref{s:het}. The translation by $\ff12e_F$ that was allowed by
discrete degrees of freedom coming from the R-R sector of the type II
string plays an essential r\^ole in finding the massless spectrum of
the theory at the most interesting point in the moduli space of theories.


\section{A Phase Transition}	\label{s:pht}

In \cite{GMS:con} an example of a phase transition was given. This
example consisted of a conifold transition in which one degenerates
the complex structure of the starting manifold to acquire a collection
of isolated ``nodes'' each of which is locally of the form
\begin{equation}
  w^2+x^2+y^2+z^2=0,	\label{eq:node}
\end{equation}
in $\C^4$ with coordinates $(w,x,y,z)$. If this is done in just the
right way then the resulting singular space can be blown-up by a small
resolution (replacing points by $\P^1$'s) into a new manifold.

Since (\ref{eq:node}) is one of simplest singularities one obtains in
three dimensions it is reasonable to expect conifold transitions to be
fairly common. There is another singularity which is just as
simple: namely
\begin{equation}
  x^2+y^2+z^2=0.	\label{eq:dblpt}
\end{equation}
In this case, the singularity occurs for all $w$ and so occurs along a
whole object of complex dimension one within the threefold. That is,
(\ref{eq:node}) is a singularity of codimension three whereas
(\ref{eq:dblpt}) is codimension two. Codimension two singularities can
appear in K3 surfaces and have been considered in
\cite{W:dyn,me:enhg}. In the K3 case one can obtain enhanced non-abelian
gauge groups when such singularities appear and then (\ref{eq:dblpt})
gives $SU(2)$. It therefore seems not
unreasonable to expect the same thing to happen for threefolds. Note
that there is no reason to expect enhanced gauge groups for conifold
points, i.e., codimension three singularities. In fact, an example of
the heterotic dual picture of a conifold transition was conjectured in
\cite{KV:N=2}\footnote{In this case the gauge group contains a
nonabelian factor of $E_6$ but this plays no essential r\^ole in the
transition. It is present both before and after the conifold
transition.} and no enhancement occurs.

In our case $X^\sharp$ contains a whole $\P^1$ of singularities. At a
generic point on this curve the singularity is of the form
(\ref{eq:dblpt}). It should not come as a surprise then to get an
enhanced $SU(2)$ at this point in the moduli space of $X$'s.

At four points on the $\P^1$ of singularities the singularity is worse
than (\ref{eq:dblpt}).
These singularities are related to degenerate Enriques surfaces of the
form discussed in section \ref{s:deg}.

Now let us try to resolve $X^\sharp$ to obtain a new
manifold. Consider first a $\Z_2$-action on a K3 surface which does
not act freely but rather fixes a rational curve. Consider the limit
of the resulting quotient as we blow this fixed rational curve down
to a point. It is not hard to convince oneself that the resulting space
is essentially the same degenerate
Enriques surface we had above. Taking this up to three complex
dimensions it is clear how to proceed. Let $Y={\rm K3}\times T^2$ and
consider a group $G^\prime\cong\Z_2$ generated by $g^\prime$ which acts
as a symmetry of $Y$. $g^\prime$ is similar to $g$ in that it has the
same action on $T^2$ and satisfies (\ref{eq:Enri}). The difference is
that $g^\prime$ fixes a rational curve, $C$, within the K3 surface. Let
$Z_0=Y/G^\prime$.

The space $Z_0$ is itself singular as the orbifold had fixed
points. We can blow $Z_0$ up to obtain $Z$ which is smooth.
On the other hand, by varying the K\"ahler form on $Z_0$ we can change
the area of
$C$. In the limit that this area goes to zero we claim that $Z_0$
becomes the same space as $X^\sharp$. This shows that $X$ can be
deformed into $Z$ by going via the singular space $X^\sharp$ as
\begin{equation}
\begin{array}{ccc}
  &&Z\\
  &&\Big\downarrow\rlap{$\pi_a$}\\
  &&Z_0\\
  &&\Big\downarrow\rlap{$\pi_b$}\\
  X&\mathop{\longrightarrow}\limits^{\hbox{\scriptsize deform}}&X^\sharp
\end{array}
\end{equation}
The maps $\pi_a$ and $\pi_b$ are both blow-downs, i.e., deformations
of K\"ahler form and the horizontal map is a deformation of complex
structure.

\iffigs
\begin{figure}
  \centerline{\epsfxsize=10cm\epsfbox{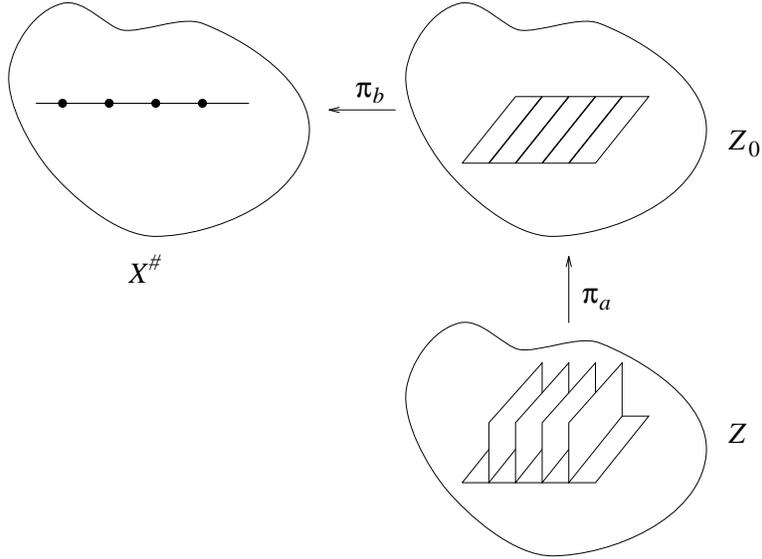}}
  \caption{Blowing up $X^\sharp$ into $Z$.}
  \label{fig:b}
\end{figure}
\fi

The Hodge numbers of $Z$ are simple to compute. The existence of a
rational curve to be fixed on the K3 surface will obstruct one
deformation of complex structure relative to $X$. Thus
$h^{2,1}(Z)=10$. The orbifold $Z_0$ has 4 fixed $\P^1$'s. Blowing these
up gives a contribution of 4 to $h^{1,1}$ from $\pi_a$. $\pi_b$ blows
just one curve down and so contributes 1 to $h^{1,1}$. Thus,
$h^{1,1}(Z)=11+4+1=16$. One can also show that $Z$ is
simply-connected. We show the blow-up schematically in figure
\ref{fig:b} by showing complex dimensions as real.

The heterotic story of this transition was given in
\cite{FHSV:N=2}. We degenerate the original theory in the ``Coulomb''
phase by varying a scalar
from a vector multiplet to enhance one of the $U(1)$ factors in the
gauge group to $SU(2)$. At this point four hypermultiplets in the form
of {\bf 2}'s of this $SU(2)$ become massless. By giving non-zero
values to the scalars in these new hypermultiplets we can venture off
into the ``Higgs'' phase. The number of extra quaternionic dimensions
from these hypermultiplets in this new
phase is $4\times{\bf 2}-\dim SU(2)=5$. The effect of this is to
completely break $SU(2)$ including any $U(1)$ subgroup. Thus there are
$11+5=16$ hypermultiplets and $11-1=10$ vector multiplets in agreement
with the type IIB picture above.

In \cite{FHSV:N=2} there was some speculation that $Z$ might be some
particular complete intersection in a weighted projective space. This
was based purely on the Hodge numbers and we will not try to verify
this conjecture here.

\begin{figure}
\setlength{\unitlength}{0.008in}%
$$\begin{picture}(445,266)(60,385)
\thinlines
\put(100,420){\circle*{6}}
\put(140,420){\circle*{6}}
\put(260,420){\circle*{6}}
\put(300,420){\circle*{6}}
\put(420,420){\circle*{6}}
\put(460,420){\circle*{6}}
\put(120,440){\circle*{6}}
\put(160,440){\circle*{6}}
\put(240,440){\circle*{6}}
\put(280,440){\circle*{6}}
\put(320,440){\circle*{6}}
\put(400,440){\circle*{6}}
\put(440,440){\circle*{6}}
\put(140,460){\circle*{6}}
\put(180,460){\circle*{6}}
\put(220,460){\circle*{6}}
\put(260,460){\circle*{6}}
\put(300,460){\circle*{6}}
\put(340,460){\circle*{6}}
\put(380,460){\circle*{6}}
\put(420,460){\circle*{6}}
\put(160,480){\circle*{6}}
\put(200,480){\circle*{6}}
\put(240,480){\circle*{6}}
\put(280,480){\circle*{6}}
\put(360,480){\circle*{6}}
\put(400,480){\circle*{6}}
\put(180,500){\circle*{6}}
\put(220,500){\circle*{6}}
\put(260,500){\circle*{6}}
\put(300,500){\circle*{6}}
\put(340,500){\circle*{6}}
\put(380,500){\circle*{6}}
\put(200,520){\circle*{6}}
\put(240,520){\circle*{6}}
\put(280,520){\circle*{6}}
\put(320,520){\circle*{6}}
\put(360,520){\circle*{6}}
\put(220,540){\circle*{6}}
\put(260,540){\circle*{6}}
\put(300,540){\circle*{6}}
\put(340,540){\circle*{6}}
\put(240,560){\circle*{6}}
\put(280,560){\circle*{6}}
\put(320,560){\circle*{6}}
\put(260,580){\circle*{6}}
\put(300,580){\circle*{6}}
\put(280,600){\circle*{6}}
\put(320,480){\circle*{6}}
\put(120,400){\circle{10}}
\put(280,400){\circle{10}}
\put(440,400){\circle{10}}
\put(200,440){\circle{10}}
\put(200,480){\circle{10}}
\put(360,440){\circle{10}}
\put(360,480){\circle{10}}
\put(280,480){\circle{10}}
\put(280,520){\circle{10}}
\put(280,560){\circle{10}}
\put(280,600){\circle{10}}
\put(280,440){\circle{10}}
\put(120,440){\circle{10}}
\put(440,440){\circle{10}}
\put( 80,400){\line( 1, 0){420}}
\put( 80,400){\line( 0, 1){240}}
\put(295,615){\framebox(10,10){}}
\put(315,595){\framebox(10,10){}}
\put(335,575){\framebox(10,10){}}
\put(375,535){\framebox(10,10){}}
\put(395,515){\framebox(10,10){}}
\put(415,495){\framebox(10,10){}}
\put(435,475){\framebox(10,10){}}
\put(455,455){\framebox(10,10){}}
\put(475,435){\framebox(10,10){}}
\put(355,515){\framebox(10,10){}}
\put(355,555){\framebox(10,10){}}
\put( 75,379){\makebox(0,0)[lb]{\raisebox{0pt}[0pt][0pt]{0}}}
\put(175,379){\makebox(0,0)[lb]{\raisebox{0pt}[0pt][0pt]{5}}}
\put(270,379){\makebox(0,0)[lb]{\raisebox{0pt}[0pt][0pt]{10}}}
\put(370,379){\makebox(0,0)[lb]{\raisebox{0pt}[0pt][0pt]{15}}}
\put(470,379){\makebox(0,0)[lb]{\raisebox{0pt}[0pt][0pt]{20}}}
\put( 65,495){\makebox(0,0)[lb]{\raisebox{0pt}[0pt][0pt]{5}}}
\put( 55,595){\makebox(0,0)[lb]{\raisebox{0pt}[0pt][0pt]{10}}}
\put(505,395){\makebox(0,0)[lb]{\raisebox{0pt}[0pt][0pt]{$r$}}}
\put( 75,645){\makebox(0,0)[lb]{\raisebox{0pt}[0pt][0pt]{$a$}}}
\put(345,620){\vector(-1,0){35}}
\put(350,615){\makebox(0,0)[lb]{\raisebox{0pt}[0pt][0pt]{$Z$}}}
\put(235,600){\vector(1,0){35}}
\put(217,595){\makebox(0,0)[lb]{\raisebox{0pt}[0pt][0pt]{$X$}}}
\end{picture}$$
	\caption{Voisin's \CY\ manifolds. ($X$ is the circle.)}
	\label{fig:V}
\end{figure}
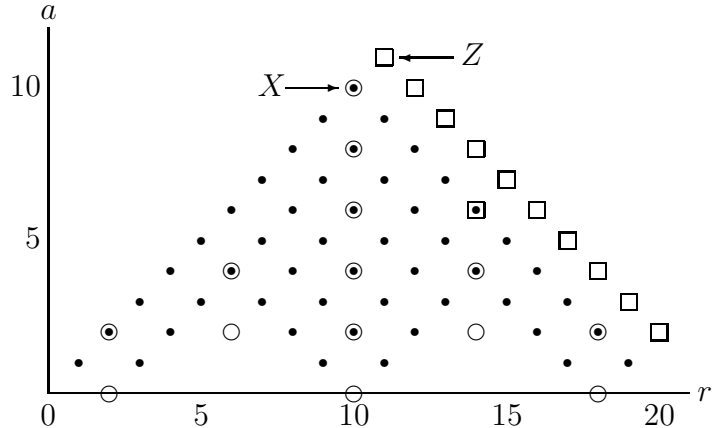

Both $X$ and $Z$ are members of the set of ``Voisin'' \CY\ manifolds
studied in \cite{Voi:K3,Bor:K3}. Consider $({\rm K3}\times T^2)/\Z_2$
where the involution acts, as before, by inversion on the $T^2$ and
satisfies (\ref{eq:Enri}) on the K3 surface. In general there are many
possibilities for the fixed-point set of the action on the K3 surface
and they were classified by Nikulin \cite{Nik:Z2}. Essentially the
fixed point set usually consists of a set of algebraic curves of which
one may have a
non-zero genus and the others are all rational. Each possibility leads to
different \CY\ manifolds. They are shown in figure \ref{fig:V}.
{}From this figure the Hodge numbers of the resulting manifolds are
\begin{equation}
\eqalign{h^{1,1} &= 5+3r-2a\cr
h^{2,1} &= 65-3r-2a,\cr}
\end{equation}
except for a couple of exceptions, one of which is $X$.

Each dot, circle or square in figure \ref{fig:V} represents a
different manifold. It was conjectured in \cite{Voi:K3} and then
proven in \cite{AM:K3m} that pairs related by a reflection in the
vertical $r=10$ line are mirror pairs --- hence the apparent symmetry
of the figure. Those without mirrors are shown as squares. See
\cite{Nik:Z2} for the difference between dots and circles.

It would be nice to give a model for the space to which the type IIA
string on $X$ goes upon the phase transition. Such a space must be the
mirror of $Z$. Unfortunately figure \ref{fig:V} shows that $Z$ is one
of the spaces which does not have a Voisin mirror. This should not
come a complete surprise for the following reason. When we go to the
interesting degeneration point in the type IIA moduli space, as we
explained earlier, the covering K3 surface must have volume of order
the Planck scale. When we go through the phase transition we lose the
modulus from the vector multiplet which gave us the ability to change
the volume. Therefore it seems reasonable that the space we go to from
the phase transition is ``stuck'' at the Planck scale. A more
precise statement of this is that the mirror of $Z$ has no \CY\ phase
and is of the type discussed in \cite{AG:gmi}.

The extremal transition we have considered, from $X$ to $Z$,
takes us from one Voisin manifold to another. One can certainly
envision a generalization of this process where we move to more Voisin
manifolds. All we need to do is to further degenerate the quotienting
so as to pick up more fixed points. Unfortunately it is not simple to
do this within the current context. The moduli space of complex
structures on $Z$ is a subset of those on $X$ and it follows that
every degeneration
of $Z$ can be understood in terms of a degeneration of the
covering K3 rather than a degeneration of the quotient. This does
not mean it cannot be done however as we have yet to explore the
hypermultiplet moduli space.


\section{Remarks}	\label{s:conc}

In this paper we have studied two aspects of type II string theory on
$X$, where $X$ which is an orbifold without fixed points of K3$\times
T^2$. Firstly there are quite subtle discrete degrees of freedom
coming from the R-R sector which must be employed to obtain modular
invariance of the dual heterotic string. The precise explanation of
this effect in terms of the type II string is not yet understood and
one might say we have no right to expect we can understand it until we
have a better
definition for string theory. What we did however was to give an
explicit geometrical interpretation of these degrees of freedom.

Secondly we examined a phase transition from the type IIB string on $X$
to another type IIB string theory on $Z$. It turned out that this
analysis was intimately connected to the first question of discrete
degrees of freedom. These degrees of freedom affect the numbers of
massless states precisely at the point of degeneration connecting the
two phases. This may give some further insight into these discrete
parameters. It is also gives further evidence that we have indeed
correctly identified a type II/heterotic dual pair.

Construction of the heterotic dual to the type IIA string on $X$ was
obtained by orbifolding the generally-believed string-string duality
in six dimensions compactified on a torus. One would have hoped that
this case would have been
easy as the orbifold action was free but we still had problems with
discrete degrees of freedom. In order to understand more dual pairs
with $N=2$ supersymmetry in four dimensions it would be nice to extend
this construction to orbifolds with fixed points. In \cite{VW:pairs}
it was shown that such a process is problematic. There it was stated
that ``orbifolding does not commute with duality''. It would be nice
to think that we can rescue orbifolding from such a fate by
understanding how it should be modified in the non-perturbative
context. The example studied in this paper might give some insight
into this problem as $Z$ is constructed from an orbifold with fixed
points.

Extremal transitions are performed by shrinking down
$S^2$'s, $S^3$'s and/or rational surfaces within a \CY\ manifold to
obtain a degenerate space. This is then followed by a reverse procedure
to give a new
manifold. Since the objects which are shrunk down are all simply connected it
appears, at first sight, that $\pi_1$ cannot change during such a
transition. If this were the case then $\pi_1$ would provide an
obstruction to connecting up the moduli space of all \CY\ manifolds
with extremal transitions. From the example studied in this paper we
see that $\pi_1$ can change. How this happens is clear from figure
\ref{fig:a}. The two spheres in this figure within $Y$ are identified
by the freely acting $\Z_2$ and so we can write a non-contractable
loop in $X$ as the image on the quotient of a line connecting one
sphere to the other. As we go to the degeneration, $X^\sharp$, the
spheres approach and so this line shrinks down to zero length, i.e.,
the loop in $X$ has contracted and $\pi_1(Z)=0$. Thus, as the $S^2$'s
shrink down, an $S^1$ also shrinks down as a side-effect. As $\pi_1$
can no longer be viewed as an obstruction, the possibility remains
that all \CY\ manifolds may be connected by extremal transitions.

Now that we have identified the phase into which the type II string on
$X$ passes it would be nice to do further analysis of the
corresponding Seiberg-Witten theory. In
\cite{KKL:limit} it was shown
that Seiberg-Witten theory could be recovered from the
$\alpha^\prime\to0$ limit of the dual pair conjectured in
\cite{KV:N=2} (see also
\cite{Bill:limit,KLT:limit,CLM:limit,AP:limit} for
discussions relevent to this). That is, the Yang-Mills theory of the
heterotic string
theory is linked to the algebraic geometry of the type II theory to
agree with \cite{SW:I}. Here such an analysis may prove even more
interesting because of the presence of the Higgs phase. One thing to
note is the fact that the five new
interesting hypermultiplet moduli are identified as blowing-up modes
within $Z$. That is, they are K\"ahler modes and are thus subject to
world-sheet instanton corrections. To remove such effects we need to
go to the
type IIA string on the mirror of $Z$. As explained above, it is
probably the case that $Z$ has no \CY\ mirror. This should not
discourage us however as one can still build a model for the moduli
space of complex structure of the mirror by analyzing it in some other
phase, such as a Landau-Ginzburg phase. It is not immediately
apparent how to do this as we have not written $Z$ in the form of a
complete intersection. Clearly this deserves further attention.


\section*{Acknowledgements}

A great deal of this work arose as a side-product from some
collaborative work with M.~Gross. I thank M.~Gross for many important
contributions. It is also a pleasure to thank S.~Chaudhuri,
I.~Dolgachev, B.~Greene,
D.~Morrison and A.~Strominger for useful conversations as well as the
``Plancksters'' at the
I.T.P., Santa Barbara program where part of this work was done under
the NSF grant PHY94-07194. The
work of the author is supported by a grant from the National Science
Foundation.


\end{document}